\documentstyle[12pt]{article}
  
\catcode`@=11
\def\secteqno{\@addtoreset{equation}{section}%
\def\theequation{\thesection.\arabic{equation}}}
\catcode`@=12
\secteqno
\newcommand{\be}{\begin{equation}}
\newcommand{\ee}{\end{equation}}
\newcommand{\bea}{\begin{eqnarray}}
\newcommand{\eea}{\end{eqnarray}}
\newcommand{\bref}[1]{(\ref{#1})}
\newcommand{\nn}{\nonumber}

\newcommand{\C}[1]{{\cal #1}}

\begin{document}
\thispagestyle{empty}
\hfill April 11, 2008

\hfill KEK-TH-1240
\vskip 20mm
\begin{center}
{\Large\bf $\kappa$ symmetric OSp(2$\mid$2) WZNW model}
\vskip 6mm
\medskip

\vskip 10mm
{\large Machiko\ Hatsuda$^{\ast\dagger}$~and~Yoji~Michishita$^\star$}

\parskip .15in
{\it $^\ast$Theory Division,\ High Energy Accelerator Research Organization (KEK),\\
\ Tsukuba,\ Ibaraki,\ 305-0801, Japan} \\{\it $^\dagger$Urawa University, Saitama \ 336-0974, Japan}\\
{\it $^\star$Department of Physics, Faculty of Education, Kagoshima University\\
Kagoshima, 890-0065, Japan
}\\
{\small e-mail:\ 
{mhatsuda@post.kek.jp~and~michishita@edu.kagoshima-u.ac.jp} }\\

\medskip
\end{center}
\vskip 10mm
\begin{abstract}
We construct a $\kappa$ symmetric WZNW model for the OSp(2$\mid$2) supergroup,
whose bosonic part is AdS$_3\times$S$^1$ space.
The field equation gives the chiral current conservation
and the right/left factorization is shown after the $\kappa$ symmetry is fixed.
The right-moving modes contain both bosons and fermions 
while the left-moving modes contain only bosons. 
\end{abstract}

\setcounter{page}{1}
\parskip=7pt
\newpage
\section{Introduction}

Superstrings in anti-de Sitter (AdS) spaces 
have important feature that those have conformal field theory duals.
Recently it has been discussed that
the heterotic string theory in AdS$_3$ space
has the dual heterotic nonlinear superconformal algebras
\cite{Lapan:2007jx}.
There are many studies on the AdS/CFT correspondence 
for the type II superstring theories 
whose worldsheet actions 
are known such as the $\sigma$ models
given in \cite{Metsaev:1998it}. 
The pp-wave limit of the worldsheet type II superstring
action allows the lightcone quantization \cite{Metsaev:2002re}.
Further generalization \cite{Frolov:2002av}
leads to the integrable property of the system
which became one of the guiding principles to 
explore the AdS/CFT correspondence.
On the other hand
few studies on the ones for a
 heterotic string have been done 
where an worldsheet
action of a heterotic string
in AdS space is not known so far.

The heterotic string
is a combination of a chiral bosonic
string and a chiral superstring.
A chiral superstring in flat space is well described by the Neveu-Schwarz-Ramond formulation,
but it is difficult to describe spacetime supersymmetry
in curved space because of the lack of the spacetime fermions.
A superstring in curved space
is well described by the
Green-Schwarz formulation, but it is difficult 
to separate chiral right/left-moving modes
for the worldsheet superconformal theory.
There are several formulations
of a chiral superstring in AdS space
where chiral spacetime fermions make both
spacetime supersymmetry and
right/left separation manifest from the beginning
such as 
the supergroup covariant $\sigma$ models  
\cite{Berkovits:1999im,Berkovits:1999zq,Hikida:2007sz,Berkovits:2000fe}.
In these formulations 
the existence of the kinetic term for those fermions
avoids the $\kappa$ symmetry, 
and the worldsheet conformal field theory technique is available
for the quantum computation.
The $\kappa$ symmetry is an inevitable ingredient
of the Green-Schwarz formulation
in which the kinetic term is made of only bosonic current bilinears,
and it is necessary to 
remove unphysical fermionic degrees of freedom.
There is an interesting observation \cite{Polyakov:2004br} that 
the $\kappa$ symmetric AdS strings are integrable
and dual field theories are at conformal fixed points.
In this paper
we also require
the $\kappa$ symmetry and we construct a  worldsheet action
for a ``heterotic" string in AdS space as a WZNW model.

The bosonic string in AdS$_3$ space was analyzed 
 by the SL(2) WZNW model 
\cite{Maldacena:2000hw} where chiral right/left separation
makes the quantum analysis possible.
In this paper we extend this AdS$_3$ bosonic string
 to an AdS$_3$ ``heterotic" string.
The chiral right/left separation of an abelian  $\sigma$ model
is resulted from the current conservation;
\bea
\epsilon^{\mu\nu}\partial_\mu J_\nu=0~~,~~
\partial_\mu J^{\mu}=0~~\to~~
\partial_+J_-=0=\partial_- J_+\label{wzw0}~~~.
\eea
For a non-abelian target space the flatness condition
contains the structure constant dependent term.
In order to obtain the chiral current conservation
the Wess-Zumino (WZ) term is added 
in such a way that it gives an extra contribution
to the current conservation 
\cite{Witten:1983tw};
\bea
\epsilon^{\mu\nu}(\partial_\mu J_\nu+J_\mu J_\nu)=0~~,~~
\partial_\mu (J^{\mu}+\epsilon^{\mu\nu}J_\nu)=0
~~\to~~
\partial_+J_-=0 ~\label{wzw}~~~.
\eea
Anti-chiral current conservation 
 may be constructed as
 $\partial_- \tilde{J}_+=0$ where
  $\tilde{J}_+$ must be a different function from $J_+$.

For a Green-Schwarz superstring
another type of  WZ term is 
required for the $\kappa$ symmetry
 \cite{Henneaux:1984mh}.
But this WZ term does not change
the current conservation equation.
For example the superstring in the AdS$_5\times$S$^5$ space
\cite{Metsaev:1998it,Roiban:2000yy}
there exist the non-abelian currents 
satisfying
 the flatness condition and the current conservation
\bea
\epsilon^{\mu\nu}(\partial_\mu J_\nu+J_\mu J_\nu)=0~~,~~
\partial_\mu J^{\mu}=0~~\to~~
\partial_+J_-=- \partial_- J_+=
-\frac{1}{2}J_{[+} J_{-]}\neq 0~~~,\label{wzw2}
\eea 
which is the criteria of the integrable system
 \cite{Bena:2003wd}.
The equations in \bref{wzw2}, which is neither 
\bref{wzw0} nor \bref{wzw}, do not give a chiral current conservation. 
The non-abelian bosonic WZ term should be
 also necessary for the type of 
equations in \bref{wzw}.
It is denoted that 
the currents $J_\pm$ in \bref{wzw2}
are the right-invariant (RI) currents rather than
the left-invariant (LI) currents
when the action is written in terms of the LI currents
 \cite{Hatsuda:2004it}.
The supercovariant derivatives, which are combination of 
the LI currents, are separated into
two chiral sectors on the  constrained surface  
satisfying the same Poisson bracket 
 as the one for the right/left sectors in the flat case 
 \cite{Hatsuda:2001xf,Hatsuda:2005te}. 
The problem how to reflect the 
chiral separation of the LI currents into 
the right/left separation of the RI currents
will reduce to the problem of the choice of a coordinate system
and the gauge fixing.
In this paper we construct the correct WZ term which guarantees the $\kappa$ symmetry and gives chiral currents conservation as \bref{wzw} 
for OSp(2$\mid$2) supergroup as a simplest nontrivial example.

The organization of this paper is the following:
in section 2 we review the orthosymplectic supergroup
and especially OSp(2$\mid$2) in detail
 which is used throughout 
this paper.
The group structure and parametrization 
related to the AdS$_3$ metric are presented.
In section 3 we propose 
a $\kappa$ symmetric OSp(2$\mid$2) WZNW action which will be
an action for a 
 ``heterotic" string in
  AdS$_3\times$S$^1$ space.
The parity (${\bf Z}_4$ symmetry), the $\kappa$ symmetry
and the field equations of the action are examined.    
The $\kappa$ symmetry variation is quite analogous to the 
AdS$_5$ superstring case \cite{Metsaev:1998it,Roiban:2000yy},
since the bosonic Sp(2) WZ term
does not contribute to the $\kappa$ transformation.
The $\kappa$ symmetry gauge fixing is necessary to
derive chiral right-moving current conervation.
This is familiar situation to the Green-Schwarz superstring in flat space
where the lightcone gauge is necessary 
for chiral separation to make
the worldsheet superconformal theory.
The possible solution of the field equation is proposed.
The right-moving mode contains both bosons and fermions,
but the left-moving mode contains only bosons.
In section 4 the flat limit of our action is examined.   
The current conservation equations reduce into 
Klein-Gordon equations representing free right/left-moving bosons.
The $\kappa$ gauge fixing condition and the $\kappa$ symmetry equation reduce
into a Dirac equation representing free right-moving fermions. 
\par
\vskip 6mm
\section{OSp supergroup}

We consider OSp($N$$\mid$2) as the simplest supergroup containing
SL(2,R)=Sp(2)  which could give a nontrivial WZ term.
OSp($N$$\mid$2) is the $3$-dimensional AdS group with 
$N$ supersymmetry or equally the $2$-dimensional
$N$ superconformal group. 
For $N$=2 its bosonic part is Sp(2)$\times$SO(2) corresponding to 
AdS$_3\times$S$^1$ space
and its fermionic part contains four supersymmetries. 
In this section we present concrete parametrization of Sp(2)
and OSp(2$\mid$2).
 Although concrete parametrization is not necessary
to examine the $\kappa$ symmetry and field equations,
it is necessary for a concrete expression of the action.

\par
\vskip 6mm
\subsection{AdS$_3$ }
In general an AdS$_d$ space is described by a coset
SO(2,$d$--$1$)/SO(1,$d$--$1$).
But for $d=3$ case the coset SO(2,2)/SO(1,2) reduces into 
SL(2)=Sp(2).
We choose a Sp(2) group element as
\bea
X=\displaystyle\frac{1}{\sqrt{1-x^2}}
\left({\bf I}+\displaystyle\sum_{m=0,1,2}x^m \gamma_m \right)
~~,~~
X^{-1}=\displaystyle\frac{1}{\sqrt{1-x^2}}
\left({\bf I}-\displaystyle\sum_{m=0,1,2}x^m \gamma_m \right)
\label{XXX}
\eea 
where the $\gamma$ matrix satisfies 
$\left\{\gamma_m,\gamma_n\right\}=2\eta_{mn}$ with 
$\eta_{mn}={\rm diag.}(-1,1,\dots,1)$.
It is noted that
$\omega_{\alpha\beta}$ is the Sp(2)-invariant
metric.
So  $\gamma_\alpha{}^\beta$ is not 
symmetric,
 but $\gamma_\alpha{}^\gamma\omega_{\gamma\beta}$
is  symmetric.
The LI one form for Sp(2) is given by
\bea
X^{-1}d X=\displaystyle\frac{1}{1-x^2}\displaystyle\sum_{m=0,1,2}
\gamma^m\left(d x_m-\epsilon_{mnl}x^nd  x^l\right)~~~.
\eea
The metric for AdS$_3$ space is obtained as 
\bea
ds^2=\frac{1}{2}{\rm tr} (X^{-1}d  X)^2
=\displaystyle\frac{1}{1-x^2}\displaystyle\sum_{m,n=0,1,2}
d  x^m\left(\eta_{mn}+
\displaystyle\frac{x_m x_n}{1-x^2}
\right)d  x^n~~~.\label{fubini}
\eea
If we generalize to   $d$-dimension, 
this form of the metric is invariant under the finite 
SO(2,$d$--1)~$\ni M_{\hat{m}\hat{n}}$
transformation 
with ${\hat{m}=(\natural,m)=(\natural,0,1,\cdots,d-1)}$ and omitting $\natural$ index
\bea
x_m \to x_m'=\displaystyle\frac{C_m+D_{m}{}^nx_n}{A+B^l x_l}~~,~~
M_{\hat{m}}{}^{\hat{n}}=\left(\begin{array}{cc}
A&B^n\\C_m&D_{m}{}^n
\end{array}\right)~~~.\label{frac}
\eea
A SO(2,$d$--1) matrix,
$M_{\hat{m}}{}^{\hat{n}}$, satisfies
\bea
(M^T)^{\hat{m}}{}_{\hat{n}}\eta^{\hat{n}\hat{l}}M_{\hat{l}}{}^{\hat{k}}
=\eta^{\hat{m}\hat{k}}~~,~~\eta^{\hat{m}\hat{n}}={\rm diag.}(-1,-1,1,\cdots,1)~~~
\eea
and in components 
\bea
\left\{\begin{array}{l}
-A^2+C_m\eta^{mn}C_n=-1\\
-B^m B^n+D_{l}{}^m\eta^{lk}D_{k}{}^n=\eta^{mn}\\
-AB^m+C_n\eta^{nl}D_{l}{}^m=0
\end{array}\right.~~~.
\eea
The coordinate $x_m$ is a ``projective coordinate" of the SO(2,$d$--1) group
realizing the AdS symmetry group by
the fractional linear transformation \bref{frac} 
as discussed in \cite{Hatsuda:2007it}.
Therefore the metric \bref{fubini} 
has the 2-dimensional conformal group invariance which is SO(2,2).

\par

\vskip 6mm
\subsection{OSp($N$${\mid}$$M$)}

In this subsection the general properties of
the orthosymplectic supergroup,
OSp($N$${\mid}$$M$),  are presented introducing our notation.
An OSp($N{\mid}M$) group element, $z$, satisfies 
\bea
(z^{T})^A{}_B\Omega^{BC} z_C{}^D=\Omega^{AD}~~,~~
\Omega_{AB}=\Omega^{AB}
=\left(\begin{array}{cc}{\bf 1}&{0}\\{0}&\omega\end{array}\right)
\eea
with 
${A,B,\cdots=(i,\alpha)=(1,\cdots,N, ~1,\cdots,M)}$.
It is denoted that $\omega$ is an anti-symmetric metric
with $\omega^2=-{\bf 1}$, 
so $\Omega^T \Omega={\bf 1}$. 
The Lie algebra elements $osp$($N$$\mid$$M$) $\ni M_{A}{}^{B}$ satisfy
\bea
&(M^T)^A{}_B\Omega^{BD}+\Omega^{AC} M_{C}{}^D=0  &\nn\\
&M_{AB}\equiv M_A{}^C\Omega_{CB}~\to~M_{ij}=-M_{ji}~,~M_{\alpha\beta}=M_{\beta\alpha}~,~M_{i\alpha}=M_{\alpha i}&~.
\eea
The Lie algebra is given by
\bea
\left[M_{AB},M_{CD}\right\}=\Omega_{[D|[A}M_{B)|C)}~~~
\eea
with a graded commutator; 
${\cal O}_{[AB)}={\cal O}_{AB}-(-)^{AB}{\cal O}_{BA}$
and
$\left[{\cal O},{\cal O}'\right\}={\cal O}{\cal O}'
-(-)^{{\cal O}{\cal O}'}{\cal O}'{\cal O}
$. 
For a group element $z$
the LI one form  is given by 
$L_A{}^B=d\sigma^{\mu}(L_\mu)_A{}^B
=d\sigma^{\mu}(z^{-1}\partial_\mu z)_{A}{}^B$
and we use the following notation
\bea
L_{AB}\equiv L_A{}^C\Omega_{CB}=
\left(\begin{array}{cc}{\bf L}_{ij}&L_{i\beta}\\
L_{j\alpha}&{\bf L}_{\alpha\beta}
\end{array}\right)~~{\rm with}~{\bf L}_{ij}=-{\bf L}_{ji}~,~
{\bf L}_{\alpha\beta}={\bf L}_{\beta\alpha}~~~.
\eea
They satisfy the following Maurer-Cartan equations:
\bea
&&\epsilon^{\mu\nu}\left[\partial_\mu
({\bf L}_\nu)_{ij}+({\bf L}_\mu)_{ik}({\bf L}_\nu)_{kj}
-(L_\mu)_{i\alpha}( L_\nu)_{j\beta}\omega^{\alpha\beta}\right]=0\nn\\
&&\epsilon^{\mu\nu}\left[\partial_\mu
({\bf L}_\nu)_{\alpha\beta}+({\bf L}_\mu)_{\alpha\gamma}({\bf L}_\nu)_{\delta\beta}
\omega^{\gamma\delta}
+(L_\mu)_{i\alpha}(L_\nu)_{i\beta}\right]=0\label{MCnm}\\
&&\epsilon^{\mu\nu}\left[\partial_\mu
({ L}_\nu)_{i\alpha}+({\bf L}_\mu)_{ik}({ L}_\nu)_{k\alpha}
-(L_\mu)_{i\beta}({\bf L}_\nu)_{\gamma\alpha}\omega^{\beta\gamma}\right]=0~~~.
\nn
\eea

For the OSp(2$\mid$2) group indices run as $i,j=1,2$ and $\alpha,\beta=1,2$
in the above equations. 
Denoting $M_{ij}=\epsilon_{ij}T$, $M_{\alpha\beta}=P_{\alpha\beta}$ and
$M_{i\alpha}=Q_{i\alpha}$,
its Lie algebra $osp(2{\mid}2)$ is given by 
\bea
&\left[P_{\alpha\beta},P_{\gamma\delta}\right]=\omega_{(\delta{\mid}(\alpha}
P_{\beta){\mid}\gamma)}~~,~~\left\{
Q_{i\alpha},Q_{j\beta}\right\}=-\delta_{ij}P_{\alpha\beta}
+\omega_{\alpha\beta}\epsilon_{ij}T&\label{osp22}\\
&\left[P_{\alpha\beta},Q_{i\gamma}\right]=-Q_{i(\alpha}\omega_{\beta)\gamma}
~~,~~
\left[T,Q_{k\alpha}\right]=-\epsilon_{kj}Q_{j\alpha}&\nn~~~.
\eea
The Maurer-Cartan equations for $osp(2{\mid}2)$ are given by
\bref{MCnm} without the second term in the first line
because ${\bf L}_{ik}{\bf L}_{kj}\to {\bf L}^2=0$.

\par
\vskip 6mm

\subsection{Left  OSp(2$\mid$2) invariant one forms }

In this section we give a concrete expression of 
the left OSp(2$\mid$2) invariant one forms.
We use linear parametrization for the OSp(2$\mid$2) matrix
instead of familiar exponential parametrization \cite{Metsaev:1998it}. 
There is an example of the linear parametrization of OSp supergroup 
\cite{Hatsuda:2003wt}, but we use different one as given below.

We parametrize  OSp(2$\mid$2) group elements as
\bea
z_A{}^B=\left(\begin{array}{cc}
{\bf I}&\theta\\-\omega \theta^{T}&{\bf I}
\end{array}\right)
\left(\begin{array}{cc}
\Upsilon^{-1/2}&{0}\\{0}&a^{-1/2} {\bf I}
\end{array}\right)
\left(\begin{array}{cc}
Y&{0}\\{0}&X
\end{array}\right)
\eea
where ${\bf I}$'s are 2$\times$2 unit matrices. 
It is convenient to introduce $\Upsilon_{ij}$'s   as
\bea
\Upsilon_{ij}&=&\delta_{ij}+\theta_i{}^\alpha\omega_{\alpha\beta}\theta_j{}^\beta\nn\\
\Upsilon^{-1}{}_{ij}&=&\delta_{ij}
-\displaystyle\frac{1}{a}\theta_i{}^\alpha\omega_{\alpha\beta}\theta_j{}^\beta\nn\\
\Upsilon^{1/2}{}_{ij}&=&\delta_{ij}
+\displaystyle\frac{1}{1+\sqrt{a}}\theta_i{}^\alpha\omega_{\alpha\beta}\theta_j{}^\beta\nn\\
\Upsilon^{-1/2}{}_{ij}&=&\delta_{ij}
-\displaystyle\frac{1}{\sqrt{a}(1+\sqrt{a})}
\theta_i{}^\alpha\omega_{\alpha\beta}\theta_j{}^\beta\nn\\
a&=&1-\displaystyle\frac{1}{2}\theta_i{}^\alpha\omega_{\alpha\beta}\theta_i{}^\beta~~~
\eea 
with $\Upsilon^n{}_{ij}\theta_j=a^n\theta_i$.
Then the OSp condition, $z^{T}\Omega z=\Omega$, leads to 
$Y^TY={\bf I}$ and $X^T\omega X=\omega$,
i.e. $Y \in$ O(2) and $X\in$ Sp(2).

The inverse of $z$ is given by
\bea
z^{-1}{}_A{}^B=
\left(\begin{array}{cc}
Y^{-1}&{0}\\{0}&X^{-1}
\end{array}\right)
\left(\begin{array}{cc}
\Upsilon^{-1/2}&{0}\\{0}&a^{-1/2} {\bf I}
\end{array}\right)
\left(\begin{array}{cc}
{\bf I}&-\theta\\\omega \theta^{T}&{\bf I}
\end{array}\right)~~~.
\eea
The LI one forms, $L_A{}^B=(z^{-1}\partial z)_A{}^B$, are given by
\bea
{\bf L}_{i}{}^{j}&=&\left(Y^{-1}\partial Y\right)_{i}{}^{j}
\nn\\
&&+Y^{-1}{}_{i}{}^{k}\left[
\displaystyle\frac{1}{4a(1+\sqrt{a})^2}\displaystyle\sum_{m=0,1,2}
\left(\theta \gamma_m\omega \partial\theta\right)
\theta_{[k} \gamma^m \omega
\theta_{l]}
+\displaystyle\frac{1}{\sqrt{a}(1+\sqrt{a})}
\theta_{[k} \omega\partial\theta_{l]}
\right]Y_{l}{}^{j}\nn\\
 {\bf L}_{\alpha}{}^{\beta}&=&(X^{-1}\partial X)_{\alpha}{}^{\beta}
+X^{-1}{}_{\alpha}{}^\gamma\left[
\displaystyle\frac{1}{2a}\displaystyle\sum_{m=0,1,2}(\gamma_m)_{\gamma}{}^{\delta}
\left(\theta_i \gamma^m \omega \partial\theta_i\right)
\right] X_{\delta}{}^{\beta}\label{LI1form}\\
L_{i}{}^\alpha&=&
\left(Y^{-1}\Upsilon^{-1/2}\right)_{ij} ~
\partial \theta_j{}^\gamma ~\displaystyle\frac{1}{\sqrt{a}}X_\gamma{}^\alpha
 \eea
with 
$\theta_{[k}\omega \gamma^m
\theta_{l]}=
\theta_{[k}{}^\alpha(\omega\gamma^m)_{\alpha\beta}
\theta_{l]}{}^\beta
$, 
$\theta_{[k} \omega\partial\theta_{l]}=
\theta_{[k}{}^\alpha \omega_{\alpha\beta}\partial\theta_{l]}{}^\beta
$ and
$\theta \omega\gamma_m \partial\theta=
\theta_k{}^\alpha (\omega\gamma_m)_{\alpha\beta}\partial\theta_k{}^\beta$. 

\par\vskip 6mm

\section{$\kappa$ symmetric OSp(2$\mid$2) WZNW action}

We consider the following action for a supersymmetric string
in the OSp(2${\mid}$2) 
background whose bosonic part is AdS$_3\times$S$^1$.
The criteria to construct an action are:
\begin{enumerate}
\item  it has (pseudo) global OSp(2$\mid$2) invariance;
\item  its bosonic Sp(2) part is the standard WZNW model;
\item  the WZ term is closed, $dH=0$; 
\item it has generalized  even parity, or equally ${\bf Z}_4$ invariance;   
\item  it has $\kappa$-symmetry invariance;
\item  its field equation gives 
the chiral right-moving current conservation.
\end{enumerate}

We propose the following action:
\bea
S&=&S_0+S_{WZ}\nn\\
S_0&=&\frac{1}{2T}\displaystyle\int d^2\sigma \sqrt{-h}h^{\mu\nu}{\rm Str}
\left[(z^{-1}\partial z)\mid_{\rm bosonic~part}\right]^2
\nn\\&=&
\frac{1}{2T}\displaystyle\int d^2\sigma \sqrt{-h}h^{\mu\nu}
\left[({\bf L}_\mu )_{ji}({\bf L}_\nu )_{ij}
-({\bf L}_\mu )_{\alpha\beta}\omega^{\beta\gamma}({\bf L}_\nu )_{\gamma\delta}\omega^{\delta\alpha}
\right]\label{S0}\\
S_{WZ}&=&\frac{k}{2}\displaystyle\int d^3\sigma H\nn\\
H&=&
\frac{1}{3}{\bf L}_{\alpha\beta}\omega^{\beta\gamma}{\bf L}_{\gamma\delta}
\omega^{\delta\epsilon}
{\bf L}_{\epsilon\phi}\omega^{\phi\alpha}
-L_{i}{}^{\alpha}{\bf L}_{\alpha\beta}L_{i}{}^{\beta}
-L_{i}{}^{\alpha}{\bf L}_{ij}\omega_{\alpha\beta}
L_{j}{}^{\beta}\label{SH}~~~.
\eea
The criteria 1-3 are guiding principles to determine the above form:
\begin{itemize}
\item{criterion 1:
the OSp(2$\mid$2) invariance is manifest up to total derivative
caused from the variation of the WZ term as usual,
since this action is written in terms of the LI one forms.
Furthermore we also impose another global  Sp(2) symmetry, 
${\bf L}_{\alpha\beta}
\to (g^T {\bf L}g)_{\alpha\beta}$ and $L_{i\alpha}
\to (L g)_{i\alpha}$ for Sp(2)$\in g$. 
This Sp(2) symmetry corresponds to a part of the AdS$_3$ isomentry
and it is not expressed by ``$z\to zg$" type transformation.  
This Sp(2) together with Sp(2)$\subset $OSp(2$\mid$2) forms
SO(2,2), the AdS$_3$ or the 2-dimensional conformal group,  
discussed in the subsection 2.1.}
\item{criterion 2:  the bosonic Sp(2) part of the action is obtained 
by setting $\theta=0$ and $Y={\bf I}$.
The survived  $X$ dependence 
is just standard WZNW model
\bea
S \to -\frac{1}{2T}\int{\rm tr}(X^{-1}\partial X)^2+\frac{k}{6}\int{\rm tr}(X^{-1}dX)^3 ~~~.
\eea
}
\item{criterion 3:
the three form $H$ is determined from the closure, $dH=0$, 
using the Maurer-Cartan equations in
\bref{MCnm} for $osp$(2$\mid$2).
It is also mentioned that $H$ in \bref{SH} can not be rewritten as 
${\rm Str} (z^{-1}dz)^3$.}
\end{itemize}
We will show that the action \bref{S0} and \bref{SH} 
satisfies the criteria 4-6 as below. 
\par
\vskip 6mm
\subsection{${\bf Z}_4$ invariance}

A super-AdS group   
is a ``generalized symmetric space" based on
the supersymmetrized parity, namely ${\bf Z}_4$ symmetry,
rather than an usual ``symmetric space" \cite{Berkovits:1999zq}.
The parity operation is given by
$\Pi({M})$ with 
 $\Pi^4({M})={M}$.
The invariant subalgebra, $\Pi({M})={M}$, 
is $u(1)\times u(1)$ which is denoted by ${\cal H}_0$.
The ${\bf Z}_4$ decomposition of the $osp(2{\mid}2)$ algebra is given by
\bea
\begin{array}{lcl}
{\cal H}_0=\left\{T_{12},~P_{12}+P_{21} \right\}&,&
{\cal H}_1=\left\{q_1\pm q_2 \right\}\\
{\cal H}_2=\left\{P_{11},~P_{22} \right\}&,&
{\cal H}_3=\left\{{q}'_1\pm {q}'_2 \right\}
\end{array}\label{z4}
\eea
where we denoted $Q_{i\alpha}=(q_i,~{q}'_i)$ with $q_i=Q_{i1}$ 
and ${q}'_i=Q_{i2}$. Each subspace satisfies the following algebra  
$\left[{\cal H}_n,{\cal H}_m\right]\subset {\cal H}_{n+m}$~(mod 4).

The ${\cal H}_n$ component of the LI currents is denoted by $j_{n}$.
The action \bref{S0} and \bref{SH} is expressed as
\bea
S_0&\sim&\displaystyle\int d^2\sigma 
\left[j_0j_0+j_2j_2\right]\nn\\
S_{WZ}&\sim&\displaystyle\int d^3\sigma
\left[j_0\wedge (j_2\wedge j_2
+ j_1\wedge j_3 )
+j_2\wedge (j_1\wedge j_1+j_3\wedge j_3)
\right]~~~.
\eea
All terms are of even parity or equivalently ${\bf Z}_4$ invariant.

Our ${\bf Z}_4$ classification \bref{z4} is not 
covariant under the other global Sp(2),
which is part of 3-dimensional AdS symmetry.
In the original classification in \cite{Berkovits:1999zq}
${\cal H}_0$  coincides with H for a coset G/H.
On the other hand  our space is not a coset space, 
so one might consider an empty ${\cal H}_0$.
However one of the three $sp(2)$ generators must be ${\cal H}_0$
in such a way that the bosonic tri-linear term
in the WZ action belongs to ${\cal H}_0$;
$j_2\wedge j_2\wedge j_2 \notin {\cal H}_0$.
We have chosen $P_{(12)}$ among $sp(2)$ generators, $P_{(\alpha\beta)}$, 
as ${\cal H}_0$.

In general the WZ term ${\cal L}_{WZ}$ has a surface term ambiguity.
The OSp(2$\mid$2) invariance and the ${\bf Z}_4$ invariance restrict
 ambiguous terms to be a form of $dj_0$.
A candidate term with 3-dimensional AdS symmetry is
$d{\bf L}_{ij}\epsilon^{ij}
=L_i{}^\alpha L_j{}^\beta\omega_{\beta\alpha}\epsilon^{ij}$.
A surface term does not effect the value of three form $H$, 
the field equation and the $\kappa$ gauge variation.
The local WZ term, 
which is a form of fermionic currents bilinears
such as
$j_1\wedge j_3$,
does not exist  for this system
except the surface term 
$L_i{}^\alpha L_j{}^\beta\omega_{\beta\alpha}\epsilon^{ij}
=d{\bf L}_{ij}\epsilon^{ij}
$.
Therefore our WZ term 
is unique up to this surface term ambiguity.

\par
\vskip 6mm
\subsection{$\kappa$ symmetry invariance}

The system has ``usual" Virasoro constraints, 
since the kinetic term of the action in \bref{S0}  contains only
bilinears of bosonic LI currents. 
In general the $\kappa$ symmetry variation of the action 
is proportional to the Virasoro constraints
so that it is cancelled by the variation of the Virasoro multiplier.
When the action is written in terms of the LI currents,
the $\kappa$ symmetry variation is 
a part of the local right transformation $z\to  z\Lambda$
in such a way that
 the parameter $\Lambda$ carries the same indices with the 
LI currents.
We will determine the $\kappa$ symmetry transformation rules
by
cancellation between the $z\to  z\Lambda$ variation of the action 
 and  the Virasoro term.

The LI one form is transformed under $z\to  z\Lambda$
 as
\bea
\delta_\Lambda
 (z^{-1}\partial_\mu z)_A{}^B&=& 
\partial_\mu\Lambda_A{}^B+(z^{-1}\partial_\mu z)_A{}^C\Lambda_C{}^B
-\Lambda_A{}^C(z^{-1}\partial_\mu z)_C{}^B~~~.\label{29}
\eea 
For a fermionic parameter $\lambda_{i\alpha}$ 
the LI one form in components are transformed as
\bea
(\delta_\lambda
 {\bf L}_\mu)_{ij}&=&-(L_\mu)_{i\alpha}\omega^{\alpha\beta}\lambda_{j\beta}
+\lambda_{i\alpha}\omega^{\alpha\beta}(L_\mu)_{j\beta}\nn\\
 (\delta_\lambda
 {\bf L}_\mu)_{\alpha\beta}&=&(L_\mu)_{i\alpha}\lambda_{i\beta}
-\lambda_{i\alpha}(L_\mu)_{i\beta}\label{dellambda}\\
 (\delta_\lambda
 { L}_\mu)_{i\alpha}&=&
\partial_\mu\lambda_{i\alpha}+
( {\bf L}_\mu)_{ij}\lambda_{j\alpha}
+\lambda_{i\beta}\omega^{\beta\gamma}({\bf L}_\mu)_{\gamma\alpha}\nn~~~.
\eea
It is convenient to introduce $\sqrt{-h}h^{\mu\nu}=
\displaystyle\frac{1}{e}e_+{}^{(\mu}e_{-}{}^{\nu)}$.
The variation of the kinetic term is
\bea
\delta_\lambda {\cal L}_{0}=
\displaystyle\frac{2}{T}\frac{1}{e}e_+{}^{(\mu}e_-{}^{\nu)}
\lambda_{i\alpha}\omega^{\alpha\beta}\left(
({\bf L}_\mu)_{\beta\gamma}\omega^{\gamma\delta}(L_\nu)_{i\delta}
-({\bf L}_\mu)_{ij}(L_\nu)_{j\beta}
\right)~~~.
\eea
The $\kappa$ variation of the WZ term is given by
\bea
\delta_\lambda{\cal L}_{WZ}
=\displaystyle\frac{k}{e}e_+{}^{[\mu}e_-{}^{\nu]}
\lambda_{i\alpha}\omega^{\alpha\beta}\left(
({\bf L}_\mu)_{\beta\gamma}\omega^{\gamma\delta}(L_\nu)_{i\delta}
-({\bf L}_\mu)_{ij}(L_\nu)_{j\beta}
\right)~~~\eea
where $\epsilon^{\mu\nu}=e_+{}^{[\mu}e_-{}^{\nu]}/e$ is used.

We consider the variation 
$\delta e_\pm{}^\mu=\varphi_{\pm +}e_-{}^\mu+\varphi_{\pm -}e_+{}^\mu$, 
and so 
$\delta e=(\varphi_{-+}+\varphi_{+-})e$.
Under this variation $(\sqrt{-h}h^{\mu\nu})$ is transformed as
\bea
\delta_{\varphi}(\sqrt{-h}h^{\mu\nu})=\displaystyle\frac{1}{e}
\left(
\varphi_{++}e_-{}^{(\mu}e_-{}^{\nu)}+
\varphi_{--}e_+{}^{(\mu}e_+{}^{\nu)}
\right)
\label{delhe}~~~,
\eea
and the transformed Virasoro term is given by
\bea
\delta_{\varphi} {\cal L}=\displaystyle\frac{1}{2T}\frac{1}{e}
(\varphi_{++}e_-{}^{(\mu}e_-{}^{\nu)}+\varphi_{--}e_+{}^{(\mu}e_+{}^{\nu)})
\left(({\bf L}_{\mu})_{ij}({\bf L}_{\nu})_{ji}
-({\bf L}_{\mu})_{\alpha}{}^{\beta}({\bf L}_{\nu})_{\beta}{}^{\alpha}
\right)~~
\eea
with
$({\bf L}_\mu)_\alpha{}^{\beta}=\omega^{\beta\gamma}({\bf L}_\mu)_{\alpha\gamma}$.
In the prefactor $e_-{}^{(\mu}e_-{}^{\nu)}$ and $e_+{}^{(\mu}e_+{}^{\nu)}
$ form an orthogonal basis.

The variation of the total Lagrangian is
\bea
\delta {\cal L}
&=&\displaystyle\frac{1}{2Te}
\left[
\varphi_{++}
\left\{({\bf L}_-)_{ij}({\bf L}_-)_{ji}-
({\bf L}_-)_{\alpha}{}^{\beta}({\bf L}_-)_\beta{}^{\alpha}\right\}
+\varphi_{--}\left\{({\bf L}_+)_{ij}({\bf L}_+)_{ji}-
({\bf L}_+)_\alpha{}^\beta({\bf L}_+)_\beta{}^\alpha
\right\}\right] \nn\\&&
+\left\{
\left(
\displaystyle\frac{2}{T}+k\right){\cal P}_+^{\mu\nu}+
\left(\displaystyle\frac{2}{T}-k\right){\cal P}_-^{\mu\nu}
\right\}
\lambda_{i\alpha}\omega^{\alpha\beta}
\left\{
-({\bf L}_{\mu})_{ij}(L_\nu)_{j\beta}-
({\bf L}_{\mu})_{\beta}{}^{\gamma}(L_\nu)_{i\gamma}
\right\}
~~~\nn\\\label{variation0wz}
\eea
with  the projection operator
$
{\cal P}_{\pm}^{\mu\nu}\equiv \displaystyle\frac{1}{e}e_\pm{}^\mu e_\mp{}^\nu
=\displaystyle\frac{1}{2}(\sqrt{-h}h^{\mu\nu}\pm\epsilon^{\mu\nu})
$
and $L_\pm=e_\pm{}^\mu L_\mu$.
For a case  $k=-\frac{2}{T}$, the $\lambda_{i\alpha}$ parameter is 
determined from the $\kappa$ symmetry invariance as
\bea
\lambda_{i\alpha}={\cal P}_{-}^{\rho\lambda}\left\{
({\bf L}_\rho)_{ij}(\kappa_\lambda)_{j\alpha}
-({\bf L}_\rho)_{\alpha}{}^{\beta}(\kappa_\lambda)_{i\beta}
\right\}
=({\bf L}_-)_{ij}(\kappa_+)_{j\alpha}
-({\bf L}_-)_{\alpha}{}^{\beta}(\kappa_+)_{i\beta}~~~.
\eea
The $\kappa$ symmetry invariance is obtained as
\bea
\delta {\cal L}&=&
\left(\displaystyle\frac{1}{2Te}
\varphi_{++}+\displaystyle\frac{2}{T}(\kappa_+)_{i\alpha}\omega^{\alpha\beta}(L_+)_{i\beta}
\right)
\left\{({\bf L}_-)_{ij}({\bf L}_-)_{ji}-
({\bf L}_-)_\alpha{}^\beta({\bf L}_-)_\beta{}^{\alpha}
\right\}\nn\\
&&+\displaystyle\frac{1}{2Te}\varphi_{--}\left\{({\bf L}_+)_{ij}({\bf L}_+)_{ji}-
({\bf L}_+)_\alpha{}^{\beta}({\bf L}_+)_\beta{}^{\alpha}
\right\}=0~~\nn\\
&\Leftrightarrow&
\varphi_{++}+4e(\kappa_+)_{i\alpha}\omega^{\alpha\beta}(L_+)_{i\beta}=\varphi_{--}=0~~~.
\eea
If a case $k=\frac{2}{T}$ is chosen instead of $k=-\frac{2}{T}$, then
the $\kappa$ symmetry invariance requires  the $\lambda_{i\alpha}$ parameter to be
$
\lambda_{i\alpha}=
({\bf L}_+)_{ij}(\kappa_-)_{j\alpha}
-({\bf L}_+)_{\alpha}{}^{\beta}(\kappa_-)_{i\beta}
$ 
and 
$
\varphi_{--}+4e(\kappa_-)_{i\alpha}\omega^{\alpha\beta}(L_-)_{i\beta}=0=\varphi_{++}
$.

The 
$\kappa$ transformation, $\delta z=z\left(\begin{array}{cc}
{0}&\lambda\\-\omega \lambda^T&{0}
\end{array}\right)$, 
is expressed 
in  components as
\bea
\delta \theta_i{}^\alpha&=&(\Upsilon^{1/2}Y\lambda X^{-1}a^{1/2})_i{}^\alpha\nn\\
&=&(\Upsilon^{1/2}Y)_i{}^j
\left(({\bf L}_-)_{jk}(\kappa_+)_{k\beta}
+({\bf L}_-)_{\beta\gamma}\omega^{\gamma\delta}(\kappa_+)_{j\delta}\right)
\left(-\omega
X^{-1}a^{1/2}\right)^{\beta\alpha}\nn\\
\left(\delta X X^{-1}\right)_\alpha{}^\beta&=&
-\displaystyle\frac{1}{2a}\displaystyle\sum_{m=0,1,2}
(\gamma_m)_\alpha{}^\beta
\left(\theta{} \gamma^m \omega \delta \theta{}\right)
\label{kappa}\\
\left(\delta Y Y^{-1}\right)_{ij}&=&
-\epsilon_{ij}
\left(
\displaystyle\frac{1}{\sqrt{a}(1+\sqrt{a})}
(\theta_k\omega\epsilon_{kl}
\delta \theta_l)+\displaystyle\frac{1}{4a(1+\sqrt{a})^2}
(\theta\gamma^m\omega\delta\theta)
(\theta_k\gamma_m\omega\epsilon_{kl}
 \theta_l)
\right)
~~~,\nn
\eea
where spinor indices are omitted; for example
$\theta \gamma^m\omega \delta\theta=
\theta_i{}^\alpha (\gamma^m\omega)_{\alpha\beta} \delta\theta_i{}^\beta
$.
If we use the parametrization \bref{XXX}
and $Y=e^{i\tau_2 y}$, then
the left hand sides of the bosonic equation are given by 
 $(\delta X X^{-1})_\alpha{}^\beta
=\displaystyle\sum_{m,n,l=0,1,2}
(\gamma^m)_\alpha{}^\beta~\delta x^n
\displaystyle\frac{\eta_{nm}-\epsilon_{nml}x^l}{1-x^2}$
and $(\delta Y Y^{-1})_{ij}=\epsilon_{ij}\delta y$
.

The Virasoro constraints and the
$\kappa$ symmetry equation
are obtained by
field equations, 
$\delta {\cal L}/\delta \varphi=0$ and
$\delta {\cal L}/\delta \lambda=0$:
\bea
&\left\{({\bf L}_+)_{ij}({\bf L}_+)_{ji}-
({\bf L}_+)_\alpha{}^{\beta}({\bf L}_+)_\beta{}^{\alpha}\right\}=0
~~,~~\left\{({\bf L}_-)_{ij}({\bf L}_-)_{ji}-
({\bf L}_-)_\alpha{}^{\beta}({\bf L}_-)_\beta{}^{\alpha}\right\}
=0&\\
&\left(
\frac{2}{T}+k\right)
\left\{
({\bf L}_{+})_{ij}(L_-)_{j\alpha}+
({\bf L}_{+})_\alpha{}^\beta(L_-)_{i\beta}
\right\}
+
\left(\frac{2}{T}-k\right)
\left\{
({\bf L}_{-})_{ij}(L_+)_{j\alpha}+
({\bf L}_{-})_\alpha{}^\beta(L_+)_{i\beta}
\right\}
=0
&\nn
\eea
Since we have chosen $k=-2/T$, the $\kappa$ symmetry equation
is reduced to
\bea
({\bf L}_{-})_{ij}(L_+)_{j\alpha}+
({\bf L}_{-})_\alpha{}^\beta(L_+)_{i\beta}
=0\label{kappaeq}\label{225}~~~.
\eea
These equations are written in terms of LI currents
and they are local equations.

\par
\vskip 6mm
\subsection{Chiral current conservations}

Now let us compute the chiral current conservations. 
It was shown that the conserved Noether currents 
reflecting the global 
symmetry are RI currents,
while the supercovariant derivatives and local constraints 
are made of LI currents \cite{Hatsuda:2004it,Hatsuda:2005te}.
So we  need to consider the infinitesimal variation
$\delta z z^{-1}$ which carries the same indices with
the RI currents in order to evaluate the current conservations. 
Under this variation the LI one form is transformed as
\bea
\delta ~(z^{-1}\partial_\mu z) =z^{-1}\partial_\mu (\delta z z^{-1}) z~~~.
\eea 
The three form is transformed as
\bea
\delta H&=&{\rm Str}\left[\epsilon^{\rho\mu\nu}
\left(z^{-1}\partial_\rho (\delta z z^{-1}) z\right)(z^{-1}\partial_\mu z) (z^{-1}\partial_\nu z) \right]
\nn\\
&=&\epsilon^{\rho\mu\nu}
\partial_\rho ~{\rm Str}\left[\delta z z^{-1}~ \partial_\mu\left\{(\partial_\nu z)   z^{-1}\right\}\right]
\label{delH}~~~
\eea
where the explicit expression of the supertrace is given by
\bea
&&\epsilon^{\rho\mu\nu}{\rm Str}\left(z^{-1}\partial_\rho(\delta z z^{-1}) z\right) (z^{-1}\partial_\mu z) (z^{-1}\partial_\nu z)\nn\\
&=&\epsilon^{\rho\mu\nu}\left(z^{-1}\partial_\rho(\delta z z^{-1}) z\right)_A{}^B(z^{-1}\partial_\mu z)_B{}^C (z^{-1}\partial_\nu z)_C{}^A(-1)^A\nn\\
&=&\epsilon^{\rho\mu\nu}\left(z^{-1}\partial_\rho(\delta z z^{-1}) z\right)_A{}^B(L_\mu )_{BC'}\Omega^{CC'} (L_\nu)_{CA'}\Omega^{AA'}(-1)^A\nn\\
&=&\epsilon^{\rho\mu\nu}\left[\left(z^{-1}\partial_\rho(\delta z z^{-1}) z\right)_{ij}\left\{-(L_\mu)_{j\alpha}\omega^{\alpha\beta}(L_\nu)_{i\beta}
\right\}
\right.
\nn\\&&
\left.
~~~~
+\left(z^{-1}\partial_\rho(\delta z z^{-1}) z \right)_{\alpha}{}^\beta
\left\{-({\bf L}_\mu)_{\beta\gamma}\omega^{\gamma\delta}
({\bf L}_\nu)_{\delta\epsilon}+(L_\mu)_{i\beta}L_{i\epsilon}
\right\}\omega^{\epsilon\alpha}
\right.
\nn\\&&
\left.
~~~~+2\left(z^{-1}\partial_\rho(\delta z z^{-1}) z\right)_{\alpha j}\left\{
({\bf L}_{\mu})_{jk}(L_\nu)_{k\delta}
-(L_\mu)_{j\delta}\omega^{\beta\gamma}({\bf L}_\nu)_{\gamma\delta}\right\}\omega^{\delta\alpha}
\right]~~~.
\eea 
The variation of the kinetic term is given by
\bea
\delta {\cal L}_0&=&
\frac{1}{T} \sqrt{-h}h^{\mu\nu}{\rm Str}
\left[(z^{-1}\partial_\mu z)\mid_{\rm bosonic~part}\left(z^{-1}\partial_\nu(\delta z z^{-1}) z\right)
\mid_{\rm bosonic~part}\right]\nn\\
&=&\frac{1}{T} \sqrt{-h}h^{\mu\nu}{\rm Str}
\left[
\partial_\mu(\delta z z^{-1}) ~\left\{\partial_\nu zz^{-1}
-z\left(
\begin{array}{cc}0&L_\nu\\L^{T}_\nu&0\end{array}
\right)\Omega^T z^{-1}\right\}\right]\nn~~~,
\eea
and the variation of the WZ term is given by
\bea
\delta{\cal L}_{WZ}&=&-\displaystyle\frac{k}{2}\epsilon^{\mu\nu}{\rm Str}
\partial_\mu(\delta z z^{-1}) (\partial_\nu z z^{-1})~~~.
\eea
Total variation is written as
\bea
\delta {\cal L}&=&
\left\{\left(\displaystyle\frac{1}{T}-\displaystyle\frac{k}{2} \right){\cal P}_+^{\mu\nu}
+\left(\displaystyle\frac{1}{T}+\displaystyle\frac{k}{2} \right){\cal P}_-^{\mu\nu}
\right\}{\rm Str}\left[
\partial_\mu(\delta z z^{-1}) ~(\partial_\nu z z^{-1})\right]\nn\\
&&-\displaystyle\frac{1}{T}({\cal P}_+^{\mu\nu}+{\cal P}_-^{\mu\nu})
{\rm Str}
\left[
\partial_\mu(\delta z z^{-1}) ~z\left(
\begin{array}{cc}0&L_\nu\\L^{T}_\nu&0\end{array}
\right)\Omega^Tz^{-1}\right]=0~~~.\label{223}
\eea

We consider a case $k=-2/T$.
If the fermionic one form contribution in the second line 
 is absent,
the variation \bref{223} reduces into the $\partial_+ (\partial_-zz^{-1})=0$
in the conformal gauge, ${\cal P}^{\mu\nu}_+=(\eta^{\mu\nu}+\epsilon^{\mu\nu})/2$.
The second line contribution is caused from the $\kappa$ symmetry invariance, 
and at the same time
 the $\kappa$ symmetry constraint ambiguity also exists. 
In this paper we find the $\kappa$ gauge fixing 
in which  the chiral current  conservation becomes manifest.
We take the lightcone gauge $v\neq 0$
for the bosonic LI current
$({\bf L}_-)_{\alpha\beta}\omega^{\beta\gamma}=
\left(\begin{array}{cc}
u&v\\s&t
\end{array}\right)$.
Using the $\kappa$ gauge symmetry in \bref{kappa} as
$\delta \theta_{i1}=v(\kappa_+)_{i2}+\cdots$,
we take the following gauge for the fermionic current  as
\bea
(L_+)_{i1}=0~~~.\label{224}
\eea
We could take the usual lightcone gauge 
$\left(
\begin{array}{cc}0&0\\1&0\end{array}\right)
\left(
\begin{array}{c}\theta_{i1}\\\theta_{i2}\end{array}\right)
=0$ 
or equivalently $\theta_{i1}=0$ which is similar to the 
temporal gauge. 
But we rather choose the gauge condition 
containing a derivative in \bref{224} which may be
similar to the Lorentz gauge. 
In this gauge the equation for the $\kappa$ symmetry 
\bref{kappaeq} is solved as $
(L_+)_{i2}=0$.
This together with the equation \bref{224} reduces into 
\bea
(L_+)_{i\alpha}=0~~~\label{plusfermi}.
\eea
This equation corresponds to the one for a
free right-moving fermion in a flat limit
as we will see in the next subsection.
Using the condition \bref{plusfermi}
the field equation is obtained 
from \bref{223} as
\bea
&
\partial_+
({{J}}_-)_A{}^B=0~~,~~
({{J}}_-)_A{}^B=\partial_- z z^{-1}-
\displaystyle\frac{1}{2}
z\left(
\begin{array}{cc}0&L_-\\L^{T}_-&0\end{array}
\right)\Omega^T z^{-1}\equiv
({\cal D}_-z)z^{-1}
&.\nn\\
\label{chiralcurrent}
\eea
It seems that the second term of ${{J}}_-$
is typical 
contribution caused from the $\kappa$ symmetry
as seen in the case of 
the AdS$_5\times$S$^5$ superstring 
 \cite{Bena:2003wd,Hatsuda:2004it,Hatsuda:2005te}.

We propose a solution of the equations
 \bref{plusfermi} and \bref{chiralcurrent} as
\bea
z=Z_{(-)}(x,y,\theta;~\sigma^-) \tilde{Z}_{(+)}(x,y;~\sigma^+)~~,~~\sigma^\pm=\sigma\pm\tau
\eea
where $Z_{(-)}$ is a function
 of both bosonic and fermionic right-moving coordinates
while $\tilde{Z}_{(+)}$ is a function of only 
bosonic left-moving coordinates such as
\bea
\tilde{Z}_{(+)}(\sigma^+)=\left(
\begin{array}{cc}
Y_{(+)}(\sigma^+)&{0}\\{0}&X_{(+)}(\sigma^+)
\end{array}\right)~~~.
\eea
It is straightforward to check the equation for 
the right-moving currents given in \bref{chiralcurrent} 
as follows.
The first term of the right-moving currents \bref{chiralcurrent} is
\bea
\partial_- z z^{-1}&=&
\partial_- Z_{(-)}Z_{(-)}^{-1}~~~.
\eea
The LI one forms are given by
\bea
L_-&=&z^{-1}\partial_- z~
=~
\tilde{Z}_{(+)}^{-1}Z_{(-)}^{-1}\partial_-
Z_{(-)}\tilde{Z}_{(+)}\nn
\\
&=&
\displaystyle\left(\begin{array}{cc}
\left(Y_{(+)}^{-1}Z_{(-)}^{-1}\partial_-
Z_{(-)}Y_{(+)}\right)_i{}^j&
\left(Y_{(+)}^{-1}Z_{(-)}^{-1}\partial_-
Z_{(-)}X_{(+)}\right)_i{}^\beta\\
\left(X_{(+)}^{-1}Z_{(-)}^{-1}\partial_-
Z_{(-)}Y_{(+)}\right)_\alpha{}^j&
\left(X_{(+)}^{-1}Z_{(-)}^{-1}\partial_-
Z_{(-)}X_{(+)}\right)_\alpha{}^\beta
\end{array}
\right)~~~.
\eea
So the second term of the  left-moving currents \bref{chiralcurrent} is
calculated as
\bea
&&z\left(
\begin{array}{cc}0&L_-\\L^{T}_-&0\end{array}
\right)\Omega^T z^{-1}\nn\\
&&~=
Z_{(-)}\left(
\begin{array}{cc}
Y_{(+)}&{0}\\{0}&X_{(+)}
\end{array}
\right)
\displaystyle\left(\begin{array}{cc}
{0}&
Y_{(+)}^{-1}Z_{(-)}^{-1}\partial_-
Z_{(-)}X_{(+)}\\
X_{(+)}^{-1}Z_{(-)}^{-1}\partial_-
Z_{(-)}Y_{(+)}&
{0}
\end{array}
\right)
\left(
\begin{array}{cc}
Y_{(+)}^{-1}&{0}\\{0}&X_{(+)}^{-1}
\end{array}
\right)
Z_{(-)}^{-1}\nn\\
&&~=
Z_{(-)}
\displaystyle\left(\begin{array}{cc}
{0}&
Z_{(-)}^{-1}\partial_-
Z_{(-)}\\
Z_{(-)}^{-1}\partial_-
Z_{(-)}&
{0}
\end{array}
\right)
Z_{(-)}^{-1}~~~.
\eea
Therefore the right-moving current, satisfying $\partial_+{{J}}_-=0$, 
is given as
\bea
({{J}}_-)_A{}^B&=&({\cal D}_-z) z^{-1}\nn\\
&=&\left[\partial_- Z_{(-)}
-\displaystyle\frac{1}{2}
Z_{(-)}
\displaystyle\left(\begin{array}{cc}
{0}&
Z_{(-)}^{-1}\partial_-
Z_{(-)}\\
Z_{(-)}^{-1}\partial_-
Z_{(-)}&
{0}
\end{array}
\right)
\right]Z_{(-)}^{-1}~~~.\label{rmc}
\eea
The left-moving current, 
satisfying $\partial_-\tilde{{J}}_+=0$,  is given by
\bea
(\tilde{{J}}_+)_A{}^B&=&z^{-1}\partial_+z\nn\\
&=&\tilde{Z}_{(+)}^{-1}\partial_+\tilde{Z}_{(+)}~=~
\left(
\begin{array}{cc}
Y_{(+)}^{-1}\partial_+ Y_{(+)}&{0}\\{0}&X_{(+)}^{-1}\partial_+X_{(+)}
\end{array}
\right)
~~~\label{lmc}
\eea
which contains only bosonic components without fermionic coordinate contribution.

\par
\vskip 6mm
\section{Flat limit}

In the flat limit the AdS$_3\times$S$^1$ space 
 becomes 3-dimensional Minkowski $\times$ 1-dimensional 
 Euclidean space.
It is obtained by the following rescaling 
\bea
&x^m \to x^m/R,~ y\to y/R, ~\theta \to \theta/\sqrt{R} &\nn\\
&
{\bf L}_{\alpha\beta} \to {\bf L}_{\alpha\beta}/R,~{\bf L}_{ij} \to {\bf L}_{ij}/R,~
{L}_{i\alpha} \to {L}_{i\alpha}/\sqrt{R}&
\eea
and taking $R\to \infty$ limit.  
The LI currents become 
\bea
({\bf L}_\mu)^m&=&
\partial_\mu x^m+\displaystyle\frac{1}{2}\theta_i{}^\alpha
(\gamma^m\omega)_{\alpha\beta}\partial_\mu\theta_i{}^\beta
\nn\\
({\bf L}_\mu)^y&=&
\partial_\mu y+\displaystyle\frac{1}{2}\theta_i{}^\alpha
\omega_{\alpha\beta}\epsilon_{ij}\partial_\mu\theta_j{}^\beta\label{flatcurrents}
\\
(L_\mu)_i{}^\alpha&=&\partial_\mu \theta_i{}^\alpha
~~~\nn
\eea
with
$({\bf L}_\mu)_{\alpha\beta}=(\gamma_m \omega)_{\alpha\beta}
({\bf L}_\mu)^m$ and $({\bf L}_\mu)_{ij}=\epsilon_{ij}
({\bf L}_\mu)^y$.
The action in \bref{S0} and \bref{SH}, 
which 
is  rescaled as $S\to S/R^2$,
becomes
\bea
S&=&S_0+S_{WZ}\nn\\
S_0&=&-\frac{1}{T}\displaystyle\int d^2\sigma \sqrt{-h}h^{\mu\nu}
\left[
({\bf L}_\mu)^m({\bf L}_\nu)_m+({\bf L}_\mu)^y({\bf L}_\nu)^y
\right]
\label{S0flat}~~~\\
S_{WZ}&=&
\frac{k}{2}\displaystyle\int d^3\sigma 
\left[-L_i{}^\alpha{\bf L}^m(\gamma_m \omega)_{\alpha\beta}L_i{}^\beta
-L_i{}^\alpha {\bf L}^y\epsilon_{ij}\omega_{\alpha\beta}L_j{}^\beta\right]
\nn\\
&=&
\frac{k}{2}\displaystyle\int d^2\sigma 
\epsilon^{\mu\nu}
\left[
\theta_i{}^\alpha
(\gamma_m\omega)_{\alpha\beta}\partial_\mu\theta_i{}^\beta
~\partial_\nu x^m
+ \theta_i{}^\alpha \omega_{\alpha\beta}\epsilon_{ij}
\partial_\mu\theta_j{}^\beta
~\partial_\nu y
\right]~~~.
\label{SHflat}~~~
\eea
The bosonic tri-linear term disappears from the WZ term.
The lack of the four fermi terms in the WZ term 
 is due to the 4-dimensional $N$=1
chiral spinor property:
This system has the following cyclic identity for the  
spinors, $\chi_i{}^\alpha,~(\phi^1)_i{}^\alpha,~(\phi^2)_i{}^\alpha$ and
$(\phi^3)_i{}^\alpha$,
\bea
&
\displaystyle\sum_{1,2,3~{\rm cyclic}}
\left[\left(\chi_i{}\gamma_m\omega\delta_{ij} \phi^1{}_j{}\right)
\left(\phi^2{}_k \gamma_m\omega \delta_{kl} \phi^3{}_l\right)
+\left(\chi_i \omega\epsilon_{ij} \phi^1{}_j\right)
\left(\phi^2{}_k \omega\epsilon_{kl} \phi^3{}_l\right)\right]=0&
\eea
where the Sp(2) spinor indices are contracted as
$\chi_i\gamma_m\omega\delta_{ij} (\phi^1)_j
=\chi_i{}^\alpha (\gamma_m\omega)_{\alpha\beta}\delta_{ij} (\phi^1)_j{}^\beta$
for example. 
After supplying  $\gamma$-matrices and restructuring the spinor
indices, 
this will be rewritten as a 4-dimensional covariant cyclic identity.

The global supersymmetry of the LI currents 
\bref{flatcurrents}
is in a familiar form
\bea
\delta \theta_i{}^\alpha=\varepsilon_i{}^\alpha~,~
\delta x^m=-\displaystyle\frac{1}{2}\varepsilon_i{}^\alpha
(\gamma^m\omega)_{\alpha\beta}\theta_i{}^\beta~,~
\delta y=-\displaystyle\frac{1}{2}\varepsilon_i{}^\alpha
\omega_{\alpha\beta}\epsilon_{ij}\theta_j{}^\beta~~~.
\eea
The kinetic term \bref{S0flat} is manifestly invariant 
and the WZ term \bref{SHflat}  is pseudo invariant 
under this supersymmetry as usual.

The $\kappa$ symmetry transformations 
of the action \bref{S0flat} and \bref{SHflat} 
for $k=-2/T$ are given by
\bea
&\delta \theta_i{}^\alpha=
-({\bf L}_-)^y\epsilon_{ij}(\kappa_+)_j{}^\alpha
-({\bf L}_-)^m(\kappa_+)_i{}^\beta(\gamma_m)_{\beta}{}^\alpha&\nn\\
&\delta x^m=-\displaystyle\frac{1}{2}\theta_i{}^\alpha
(\gamma^m\omega)_{\alpha\beta}\delta\theta_i{}^\beta~,~
\delta y=-\displaystyle\frac{1}{2}\theta_i{}^\alpha
\omega_{\alpha\beta}\epsilon_{ij}\delta\theta_j{}^\beta&~\label{flatkappa}\\
&\delta \sqrt{-h}h^{\mu\nu}=4\partial_+\theta_i{}^\alpha \omega_{\alpha\beta}(\kappa_+)_i{}^\beta
~e_-{}^{(\mu}e_-{}^{\nu)}
&~~.\nn
\eea

The Virasoro constraints, field equations 
for bosons and the $\kappa$ symmetry equation
are given as
\bea
&({\bf L}_+)^m({\bf L}_+)_m+({\bf L}_+)^y({\bf L}_+)^y=0~~,~~
({\bf L}_-)^m({\bf L}_-)_m+({\bf L}_-)^y({\bf L}_-)^y=0
&\nn\\
&\partial_\mu\left(
\displaystyle\frac{2}{T}\sqrt{-h}h^{\mu\nu}({\bf L}_\nu)^m
+\displaystyle\frac{k}{2}\epsilon^{\mu\nu}\theta_i^\alpha(\gamma^m\omega)_{\alpha\beta}
\partial_\nu \theta_i{}^\beta
\right)=0&\nn\\
&\partial_\mu\left(
\displaystyle\frac{2}{T}\sqrt{-h}h^{\mu\nu}({\bf L}_\nu)^y
+\displaystyle\frac{k}{2}\epsilon^{\mu\nu}\theta_i^\alpha\omega_{\alpha\beta}
\epsilon_{ij}
\partial_\nu \theta_j{}^\beta
\right)=0&\nn\\
&\left\{
(\displaystyle\frac{1}{T}+\displaystyle\frac{k}{2}){\cal P}_{+}^{\mu\nu}
+(\displaystyle\frac{1}{T}-\displaystyle\frac{k}{2}){\cal P}_{-}^{\mu\nu}
\right\}
\left\{
({\bf L}_{\mu})^m
~(\gamma_m\omega)_{\alpha\beta}\partial_\nu \theta_i{}^\beta
+({\bf L}_{\mu})^y
~\omega_{\alpha\beta}\epsilon_{ij}\partial_\nu \theta_j{}^\beta
\right\}=0&~.\nn\\\label{flateq}
\eea
For a case  $k=-\frac{2}{T}$ the $\kappa$ symmetry
 equation in \bref{flateq}
is
\bea
({\bf L}_-)^m(\gamma_m)_\alpha{}^\beta\partial_+ \theta_i{}_\beta
+({\bf L}_{-})^y
\epsilon_{ij}\partial_+ \theta_j{}_\alpha~=0~~~.\label{flatkappaeq}
\eea

Now we take the following gauge: 
The condition for the bosonic current is
$({\bf L}_-)^m(\gamma_m)_1{}^2=-v\neq 0
$.
The condition for 
the fermionic current is
  $(L_+)_{i1}=\partial_+\theta_{i1}=0$ using the $\kappa$ symmetry degree of freedom 
$\delta \theta_{i1}=v(\kappa_+)_{i}{}_2+\cdots$.
Our gauge condition in the flat limit is 
necessary condition for the
conventional lightcone gauge  as
\bea
x_0+x_1=v\tau+{\rm const.}~~,~~
\left(\begin{array}{cc}0&0\\1&0\end{array}\right)
\left(\begin{array}{c}\theta_{i1}\\\theta_{i2}\end{array}\right)=0
~~~.
\eea
The $\kappa$ equation \bref{flatkappaeq} 
leads to $(L_+)_{i2}=\partial_+\theta_{i2}=0$.
This together with the gauge fixing condition reduces into  that
the fermion coordinates satisfy
 the free right-moving Dirac equation,
\bea
(L_+)_{i\alpha}=
\partial_+ \theta_{i\alpha}=0~~\to~~\theta_{i\alpha}(\sigma^-)~~~.
\eea

The chiral 
right-moving current
in \bref{chiralcurrent}, 
satisfying $\partial_+({{J}}_-)_A{}^B=0$,
is given by
\bea
({{J}}_-)_A{}^B=\left(
\begin{array}{cc}
\epsilon_{ij}\partial_- y&\displaystyle\frac{1}{2}\partial_-\theta_i{}^\beta\\
-\displaystyle\frac{1}{2}\omega_{\alpha\gamma}\partial_-\theta_j{}^\gamma&
(\gamma_m)_\alpha{}^\beta \partial_-x^m
\end{array}
\right)~~~.
\eea 
The left-moving current 
$\tilde{J}_+=z^{-1}\partial_+ z$ is given by
\bea
(\tilde{{J}}_+)_A{}^B=\left(
\begin{array}{cc}
\epsilon_{ij}\partial_+ y&0\\0&
(\gamma_m)_\alpha{}^\beta \partial_+x^m
\end{array}
\right)~~~
\eea
 with the gauge 
$\partial_+\theta=0$.
It satisfies 
$\partial_-(\tilde{{J}}_+)_A{}^B=0$.
 The bosonic variables $x^m$ and $y$ 
 satisfy free Klein-Gordon equations for 
both right/left-moving modes
\bea
\partial_+\partial_- x^m=0=\partial_+\partial_- y~\to~
x^m=x^m(\sigma^+)+x^m(\sigma^-)~~,~~y=y(\sigma^+)+y(\sigma^-)~~~.
\eea 
Threfore our model in the flat limit 
is a 4-dimensional ``heterotic" string with the $N$=1 supersymmetric
right-moving sector and the bosonic left-moving sector.

The heterotic Green-Schwarz action in  flat space is given by 
the sum of the usual type I Green-Schwarz action plus the chiral 
current constraint term 
as shown in the original paper \cite{Gross:1985fr}.
On the other hand the chiral current conditions, \bref{224} and
 \bref{plusfermi}, 
are result of the $\kappa$-symmetry gauge in our model.
From the fact that
 our model keeps the chiral structure after the flat limit,
the $\kappa$-symmetry may be essential for the chiral separation
rather than the non-abelian bosonic WZ term.
Although the relation between different treatments of the chiral condition 
is unclear at this stage,
our model in the flat space limit corresponds to the 4-dimensional part of the
usual critical heterotic string in flat space. 

\par
\vskip 6mm
\section{Conclusion and discussions}

We have proposed a $\kappa$ symmetric WZNW model 
for OSp(2$\mid$2) supergroup.
The kinetic term contains only bosonic current bilinears
without fermionic current bilinear.
The action contains both 
the WZ term for the Sp(2) WZNW  and the one for the
Green-Schwarz superstring. 
Then we have constructed the chiral non-abelian currents 
 corresponding to the equation \bref{wzw} in the introduction.  
The non-abelian bosonic Wess-Zumino term 
does not affect the $\kappa$ symmetry 
and $\kappa$ symmetry transformation rules 
are similar to the  AdS superstring case.
It is essential that 
the $\kappa$ symmetry and the chiral current conservation
are consistent only for the same  coefficient of the WZ term.
 We have chosen the lightcone gauge 
$({\bf L}_{\rm AdS;-})_{1}{}^2\neq 0$ 
for bosonic coordinates
and the $\kappa$ gauge
$(L_+)_{i1}=0$, then the fermionic field equation gives
$(L_+)_{i2}=0$ in this gauge.
The $\kappa$ gauge condition,  $(L_+)_{i1}=0$,
contains derivative operator $\partial_+$ which
 corresponds to the Lorentz  gauge rather than
the temporal gauge.
It is a necessary condition for the usual lightcone gauge 
in the flat limit.
This  allows us to derive 
the chiral right-moving currents 
for all $osp$(2$\mid$2) components.
The right-moving current, $({{J}}_-)_A{}^B=({\cal D}_-z) z^{-1}$, satisfying
$\partial_+({{J}}_-)_A{}^B=0$ is derived from the field equation 
as \bref{chiralcurrent}.
The left-moving current, $(\tilde{{J}}_+)_A{}^B=z^{-1}\partial_+ z$, satisfying
$\partial_-(\tilde{{J}}_+)_A{}^B=0$ is obtained as \bref{lmc} 
from the factorization solution.  
The factorization is given as $z=Z_{(-)}(\sigma^-) \tilde{Z}_{(+)}(\sigma^+)$,
where $Z_{(-)}$ is a function of both bosonic and fermionic
right-moving coordinates $x,~y,~\theta$ while
$\tilde{Z}_{(+)}$ is a function of only bosonic left-moving 
coordinate  $x,~y$.
Therefore this model describes a heterotic string 
propagating in the AdS$_3\times$S$^1$ space.

This system itself is not critical and it is expected to be embedded into some larger system to describe critical string.
We have obtained the right-moving current, $(J_-)_A{}^B=({\cal D}_-z)z^{-1}$, 
which is ``right-invariant" (RI) reflecting 
the global OSp(2$\mid$2) invariance of the action.
It was shown that the RI currents 
satisfy the Poisson bracket for the AdS superalgebra  \cite{Hatsuda:2004it}
and the stress-energy tensor
is given by the supertrace of the square of the RI currents 
\cite{Hatsuda:2005te}
when the AdS superstring action is written in terms of the LI currents. 
So it may be expected that the right-moving 
RI current constructs Sugawara form giving the central charge $c_R=0$
because the dimension of $osp(2{\mid}2)$ is zero.
The left-moving current 
has only bosonic parts giving the central charge   $c_L=1+
\frac{3k}{k-2}$. 
Since the right-moving sector has the $\kappa$-symmetry invariance in addition to the 
reparametrization invariance,
the central charge has contributions from the reparametrization ghost 
and the $\kappa$-symmetry ghost whose contribution is unknown so far.
If a (chiral) superstring in curved space with the $\kappa$ symmetry
exists, there will exist a critial theory with the $\kappa$ ghost.
Our model may also describe type II string with right/left asymmetry,
so that it consistently
describes AdS$_3\times$S$^1$ part embedded into 
some larger critical system such as
AdS$_3\times$T$^4\times$S$^3$(=AdS$_3\times$S$^1\times$T$^3\times$S$^3$).
It is interesting to find more systems into which our model can be embedded, and supergravity solutions corresponding to them.


\section*{Acknowledgments}

M.H. is supported by the Grant-in-Aid for Scientific Research No. 18540287.


\end{document}